\documentclass[journal]{IEEEtran}

\usepackage{fancyhdr}
\usepackage{amsmath}
\usepackage{amssymb}
\usepackage{graphicx}
\usepackage{booktabs}
\usepackage{amsfonts}
\usepackage{multirow}
\usepackage{algorithm}
\usepackage{algorithmic}
\usepackage{mathrsfs}
\usepackage{bm}
\usepackage{exscale}
\usepackage{relsize}
\usepackage{setspace}
\usepackage{color}
\usepackage{ulem}
\usepackage{float}

\ifCLASSINFOpdf
\else
\fi

\begin{document}

\title{Spectral-Efficient Analog Precoding for Generalized Spatial Modulation Aided MmWave MIMO}

\author{
Longzhuang~He,~\IEEEmembership{Student~Member,~IEEE},
Jintao~Wang,~\IEEEmembership{Senior~Member,~IEEE},
and Jian~Song,~\IEEEmembership{Fellow,~IEEE}

\thanks{
Longzhuang He, Jintao Wang and Jian Song are with the Department of Electronic Engineering, Tsinghua University, Beijing, 100084, China (e-mail: helongzhuang@126.com; \{wangjintao, jsong\}@tsinghua.edu.cn).

This work was supported by the National Natural Science Foundation of China (Grant No. 61471221 and No. 61471219).
}
}

% make the title area
\maketitle
\begin{abstract}
Generalized spatial modulation (GenSM) aided millimeter wave (mmWave) multiple-input multiple-output (MIMO) has recently received substantial academic attention. However, due to the insufficient exploitation of the transmitter's knowledge of the channel state information (CSI), the achievable rates of state-of-the-art GenSM-aided mmWave MIMO systems are far from being optimal. Against this background, a novel analog precoding scheme is proposed in this paper to improve the spectral efficiency (SE) of conventional GenSM-aided mmWave MIMOs. More specifically, we firstly manage to lower-bound the achievable SE of GenSM-aided mmWave MIMO with a closed-form expression. Secondly, by exploiting this lower bound as a cost function, a low-complexity iterative algorithm is proposed to design the analog precoder for SE maximization. Finally, numerical simulations are conducted to substantiate the superior performance of the proposed design with respect to state-of-the-art GenSM-aided mmWave MIMO schemes.
\end{abstract}

\IEEEpeerreviewmaketitle

\section{Introduction}
Millimeter wave (mmWave) multiple-input multiple-output (MIMO) has been proved an effective technique to substantially improve the data rates for 5G telecommunication networks \cite{rangan2014mmWave}. On the one hand, the available bandwidth provided by the mmWave frequency band is orders of magnitude larger than that provided by the current cellular communication operated in microwave bands. On the other hand, the incorporation of MIMO precoding has also facilitated a significant improvement of the spectral efficiency (SE) achieved by mmWave systems.

The concept of generalized spatial modulation (GenSM) is another MIMO technique that is recently proposed to reduce the number of radio frequency (RF) chains \cite{younis2010generalised}\cite{wang2012generalised}. In GenSM systems, only a subset of antennas are randomly activated to transmit the classic amplitude-phase modulation (APM) symbols, and the information is thus carried by the indices of the active antennas as well as the transmitted APM symbols.

To explicitly benefit from the reduced-RF-chain nature of GenSM, mmWave MIMO has been recently combined with GenSM to yield the concept of GenSM-aided mmWave MIMO in \cite{liu2015SSK}-\cite{ishikawa2016GSM}. More specifically, in \cite{liu2015SSK}, the application of space shift keying (SSK) \cite{jeganathan2009space} in indoor line-of-sight (LoS) mmWave channels was investigated, in which the authors proposed to elaborately design the antenna alignment for performance optimization. The study of \cite{liu2015SSK} was further generalized to the case of GenSM in \cite{liu2016LOS}. However, the issue of transmit precoding was not accounted for in \cite{liu2015SSK} and \cite{liu2016LOS}. In \cite{ishikawa2016GSM}, a novel GenSM-aided mmWave MIMO scheme was proposed, in which an array of phase shifters (PSs) were allocated to the antenna elements to perform analog precoding. However, the analog precoder of \cite{ishikawa2016GSM} failed to fully exploit the transmitter's knowledge of the channel state information (CSI), hence the scheme of \cite{ishikawa2016GSM} suffered from severe performance loss, especially in fading channels.

In this context, the major contributions of this paper are summarized as follows.

\begin{itemize}
  \item A GenSM-aided mmWave MIMO scheme is proposed, of which the achievable SE is lower-bounded with a closed-form expression. Aided with a constant shift, the proposed SE expression is shown to provide an accurate approximation to the true SE.

  \item By exploiting the proposed lower bound as a cost function, we propose a novel low-complexity iterative algorithm to design the analog precoder with respect to SE maximization. Finally, the achievable SE of the proposed scheme is shown by numerical simulations to significantly outperform the performance achieved by state-of-the-art GenSM-aided mmWave MIMO schemes.
\end{itemize}

The remainder of this paper is organized as follows. Section \uppercase\expandafter{\romannumeral2} introduces the system model of our proposed GenSM-aided mmWave MIMO along with the theoretical SE analysis. The proposed precoder design algorithm is introduced in Section \uppercase\expandafter{\romannumeral3}. Section \uppercase\expandafter{\romannumeral4} presents the numerical simulation results, and Section \uppercase\expandafter{\romannumeral5} concludes this paper.

\textit{Notations}: $\mathcal{CN}(\bm\mu, \bm\Sigma)$ is a circularly symmetric complex-valued multi-variate Gaussian distribution with mean $\bm\mu$ and covariance $\bm\Sigma$, while $\mathcal{CN}(\mathbf{x}; \bm\mu, \bm\Sigma)$ denotes the probability density function (PDF) of a random vector $\mathbf{x} \sim \mathcal{CN}(\bm\mu, \bm\Sigma)$. $\mathbf{M}_{(i, j)}$ is used to denote the $(i; j)$ component of a matrix $\mathbf{M}$, and $\vert \mathbf{M} \vert$ represents the determinant. $\mathbf{I}_N$ denotes an $N$-dimensional identity matrix.

\section{System Model and Spectral Efficiency Analysis}
\subsection{System Model}

\begin{figure}
\center{\includegraphics[width=0.80\linewidth]{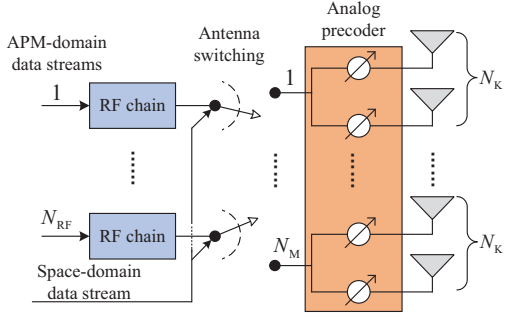}}
\caption{Block diagram of the proposed GenSM-aided mmWave MIMO scheme with analog precoding.}
\label{Fig_SystemModel}
\end{figure}

We consider an mmWave MIMO scheme with $N_\text{T}$ transmit antennas (TAs) and $N_\text{R}$ receive antennas (RAs). As depicted in Fig.\ref{Fig_SystemModel}, the $N_\text{T}$ TAs are divided into $N_\text{M}$ antenna groups (AGs), and each AG is composed of $N_\text{K}$ antenna elements, which yields $N_\text{T} = N_\text{M} N_\text{K}$. To incorporate the principle of GenSM, in this paper we assume $N_\text{M} \ge N_\text{RF}$, where $N_\text{RF}$ denotes the number of RF chains. The space-domain data stream is thus designed to randomly assign the outputs of the $N_\text{RF}$ RF chains to $N_\text{RF}$ out of the $N_\text{M}$ AGs, which is referred to as a combination of active AGs (AGC), while the rest unassigned $\left( N_\text{M} - N_\text{RF} \right)$ AGs remain inactive during this symbol's transmission period. In each symbol period, one out of the $M$ AGCs is selected by the space-domain information, and $M$ is configured as in \cite{wang2012generalised}:
\begin{equation}
  M = 2^{\left\lfloor \log_2 \binom{N_\text{M}}{N_\text{RF}} \right\rfloor},
  \label{GenSM_M_Definition}
\end{equation}
where $\lfloor \cdot \rfloor$ represents the floor operation and $\binom{\cdot}{\cdot}$ represents the binomial coefficient. To characterize this random antenna-switching regime, we use $\mathbf{u}_m \triangleq [u_{m1}, \ldots, u_{mN_\text{RF}}]^T$ to denote the indices of the AGs selected by the $m$-th AGC ($m=1, 2, \ldots, M$), which facilitates the formulation of \textit{AG-selection matrix} $\mathbf{C}_m \in \mathbb{R}_{N_\text{T} \times N_\text{RF}}$ as follows:
\begin{equation}
  [\mathbf{C}_m]_{\left[(u_{mj}-1)N_\text{K}+1 : u_{mj}N_\text{K}, \,\, j\right]} = \mathbf{1}_{N_\text{K}}, \,\, 1 \le j \le N_\text{RF},
  \label{AGSelMat}
\end{equation}
where $\mathbf{1}_{N_\text{K}} \in \mathbb{R}_{N_\text{K} \times 1}$ is an $N_\text{K}$-dimensional all-one vector, and $[\mathbf{C}_m]_{(k:n, \,\, j)}$ represents the $k$-th to $n$-th components on the $j$-th column of $\mathbf{C}_m$. Note that the rest unspecified components of $\mathbf{C}_m$ are all zeros.

Similar to \cite{ishikawa2016GSM}, $N_\text{K}$ PSs are allocated in each AG to perform analog precoding, hence we can use a diagonal matrix $\mathbf{A} \in \mathbb{C}_{N_\text{T} \times N_\text{T}}$ to represent the \textit{analog precoding matrix}, which is given by:
\begin{equation}
  \mathbf{A} \triangleq \displaystyle \text{diag}\left( \frac{1}{\sqrt{N_\text{K}}} e^{j\psi_1}, \frac{1}{\sqrt{N_\text{K}}} e^{j\psi_2}, \ldots, \frac{1}{\sqrt{N_\text{K}}} e^{j\psi_{N_\text{T}}}  \right),
  \label{A_Form}
\end{equation}
where $\psi_{n} \in [-\pi, \pi)$ denotes the rotation angle of the $n$-th TA. Finally, based on (\ref{AGSelMat}) and (\ref{A_Form}), the received signal vector $\mathbf{y} \in \mathbb{C}_{N_\text{R} \times 1}$ is given as follows when the $m$-th AGC is selected:
\begin{equation}
  \mathbf{y} = \sqrt{\rho} \mathbf{H}\mathbf{A}\mathbf{C}_m \mathbf{x} + \mathbf{n},
  \label{ReceivedSignal}
\end{equation}
where $\mathbf{H} \in \mathbb{C}_{N_\text{R} \times N_\text{T}}$ is the narrowband channel matrix, which is normalized so that $E\left\{\|\mathbf{H}\|_F^2\right\} = N_\text{R} N_\text{T}$ \cite{ayach2014spatially}. The transmitted APM symbol vector is $\mathbf{x} \in \mathbb{C}_{N_\text{RF}\times 1}$, which is distributed as $\mathbf{x} \sim \mathcal{CN}(\mathbf{0}, \mathbf{I}_{N_\text{RF}} / N_\text{RF})$. The average transmit power is given by $\rho$, while $\mathbf{n} \sim \mathcal{CN}(\mathbf{0}, \sigma_\text{N}^2 \mathbf{I}_{N_\text{R}})$ is the additive white Gaussian noise (AWGN) vector.

In this paper, we adopt the narrowband Saleh-Valenzuela channel model to characterize the insufficient-scattering and low-rank nature of mmWave signal \cite{gao2016energy}\cite{ayach2014spatially}, i.e.
\begin{equation}
  \mathbf{H} = \gamma \sum_{l=1}^L \alpha_l \Lambda_\text{t}(\phi_l^\text{t}, \theta_l^\text{t}) \Lambda_\text{r}(\phi_l^\text{r}, \theta_l^\text{r}) \mathbf{b}_\text{t}(\phi_l^\text{t}, \theta_l^\text{t}) \mathbf{b}_\text{r}^H(\phi_l^\text{r}, \theta_l^\text{r}),
  \label{SVChannl}
\end{equation}
where $\gamma = \sqrt{N_\text{T} N_\text{R} / L}$ is the normalizing factor and $L$ represents the total number of effective scattering paths. For the purpose of brevity, more details on the specific distribution of the parameters in channel model (\ref{SVChannl}) can be referred to \cite{gao2016energy}. It is worth noting that, the channel in (\ref{SVChannl}) is not the only model suitable for the algorithms and analysis in this paper. Some alternative channel models, such as \cite{samimi20163d}, are also applicable, since the analysis in this paper is only relevant to the instantaneous channel realization $\mathbf{H}$.

\subsection{Theoretical SE Analysis}
Based on Equation (\ref{ReceivedSignal}), the achievable SE of the proposed scheme can be quantified via the mutual information (MI) between $\mathbf{y}$, $\mathbf{x}$ and $m$, i.e.
\begin{equation}
  R(\mathbf{H}, \mathbf{A}) = I(\mathbf{y}; \mathbf{x}, m) = I(\mathbf{y}; \mathbf{x} \vert m) + I(\mathbf{y}; m),
  \label{RTrue}
\end{equation}
where $I(\mathbf{y}; \mathbf{x} \vert m)$ represents the APM-domain transmitted MI given that the selected AGC is known by the receiver, and can be quantified using Shannon's continuous-input continuous-output memoryless channel's (CCMC) capacity \cite{ayach2014spatially}, i.e.
\begin{equation}
  \displaystyle I(\mathbf{y}; \mathbf{x} \vert m) = \frac{1}{M} \sum_{m=1}^M \log_2 \left(\left| \frac{1}{\sigma_\text{N}^2} \bm\Sigma_m \right|\right),
  \label{RTrue_APM}
\end{equation}
where $\bm\Sigma_m \in \mathbb{C}_{N_\text{R}\times N_\text{R}}$ is given as follows:
\begin{equation}
  \bm\Sigma_m \triangleq \displaystyle \sigma_\text{N}^2\mathbf{I}_{N_\text{R}} + \frac{\rho}{N_\text{S}} \mathbf{HAC}_m \mathbf{C}_m^H \mathbf{A}^H \mathbf{H}^H.
  \label{Sigma_m_Definition}
\end{equation}

The space-domain MI term $I(\mathbf{y}; m)$ represents the information conveyed by the random antenna-switching. As the conditional distribution of $\mathbf{y}$ (given that the $m$-th AGC is selected) is $\mathcal{P}(\mathbf{y} \vert m) = \mathcal{CN}(\mathbf{y}; \mathbf{0}, \bm\Sigma_m)$ according to (\ref{ReceivedSignal}), the space-domain MI term $I(\mathbf{y}; m)$ can thus be derived as follows:
\begin{equation}
  I(\mathbf{y}; m) = \displaystyle \frac{1}{M} \sum_{n=1}^M \left( \Xi_{n1} - \Xi_{n2} \right),
  \label{RTrue_Space}
\end{equation}
where $\Xi_{n1}$ and $\Xi_{n2}$ are given as:

\begin{equation}
\arraycolsep=1.0pt\def\arraystretch{1.5}
  \begin{array}{rcl}
  \Xi_{n1} &=&   \int \mathcal{P}(\mathbf{y} \vert n) \log_2 \mathcal{P}(\mathbf{y} \vert n) \text{d} \mathbf{y} \overset{\text{(a)}}{=} \displaystyle - \log_2 \left( \left| \pi e \bm\Sigma_n \right| \right), \\
  \Xi_{n2} &=&   \int \mathcal{P}(\mathbf{y} \vert n) \log_2 \left[ \frac{1}{M} \sum_{t=1}^M \mathcal{P}(\mathbf{y} \vert t)\right] \text{d} \mathbf{y} \\
  &\overset{\text{(b)}}{\le} &   -N_\text{R} \log_2 \pi + \log_2 \left[ \sum_{t=1}^M \frac{\left| \bm\Sigma_t + \bm\Sigma_n \right|^{-1}}{M} \right],
  \end{array}
  \label{RTrue_SpaceX}
\end{equation}
in which (a) is obtained by invoking the expression of $\mathcal{P}(\mathbf{y} \vert m)$, while (b) is obtained by exploiting the concavity of $\log_2(\cdot)$ and applying Jensen's inequality. Substituting (\ref{RTrue_SpaceX}) into (\ref{RTrue_Space}), a closed-form lower bound for $I(\mathbf{y}; m)$ can thus be yielded as:
\begin{equation}
  I(\mathbf{y}; m) \ge \displaystyle \log_2 \frac{M}{e^{N_\text{R}}} - \frac{1}{M} \sum_{n=1}^M \log_2 \left( \sum_{t=1}^M \frac{\left| \bm\Sigma_n \right|}{\left| \bm\Sigma_n + \bm\Sigma_t \right|} \right).
  \label{RLB_Space_Unshift}
\end{equation}

Note that $I(\mathbf{y}; m)$ is yielded by a finite-size alphabet, i.e. $I(\mathbf{y}; m)=0$ and $I(\mathbf{y}; m)=\log_2 M$ hold at an asymptotically low and high SNR, respectively. However, it can be easily derived that the right-hand side of (\ref{RLB_Space_Unshift}) equals $N_\text{R}\log_2 \frac{2}{e}$ and $\log_2 M + N_\text{R}\log_2 \frac{2}{e}$ at low and high SNR regions. In order to obtain a more accurate approximation, we compensate this asymptotic bias, i.e. $N_\text{R}\log_2 \frac{2}{e}$ for $I(\mathbf{y}; m)$, which yields the following closed-form MI approximation to $R(\mathbf{H}, \mathbf{A})$:
\begin{equation}
  R_\text{CF}(\mathbf{H}, \mathbf{A}) = -\displaystyle \frac{1}{M} \sum_{n=1}^M \log_2 \left( \sum_{t=1}^M \frac{ \left( 2\sigma_\text{N}^2 \right)^{N_\text{R}} / M }{\left| \bm\Sigma_n + \bm\Sigma_t \right|} \right),
  \label{RLB}
\end{equation}
which is yielded by taking the expressions of (\ref{RTrue_APM}), (\ref{RLB_Space_Unshift}) into (\ref{RTrue}) and adding a constant shift $N_\text{R} \log_2 \frac{2}{e}$. Note that the subscript ``$\text{CF}$'' represents ``closed form''.

\subsection{Approximation Accuracy}
\begin{table}
\small
\caption{Simulation Parameters}
\newcommand{\tabincell}[2]{\begin{tabular}{@{}#1@{}}#2\end{tabular}}
\centering
\renewcommand\arraystretch{1.2}
\begin{tabular}{c|l|l}
\hline\hline
Symbols                 & Specifications & Typical Values \\\hline
$N_\text{T}$            & Number of TAs                                             & $8$ \\\hline
$N_\text{R}$            & Number of RAs                                             & $8$ \\\hline
$N_\text{K}$            & \tabincell{l}{Number of TAs \\ in each antenna group}                       & $2$ \\\hline
$N_\text{M}$            & Number of antenna groups                                  & $4$ \\\hline
$N_\text{RF}$           & Number of RF chains                                       & $2$ \\\hline
$\rho / \sigma_\text{N}^2$  & SNR                                                   & $0$ dB \\\hline
\hline
\end{tabular}
\label{TABLESimu}
\end{table}

We now provide numerical simulations to validate the accuracy of the proposed $R_\text{CF}(\mathbf{H}, \mathbf{A})$ as an approximation to $R(\mathbf{H}, \mathbf{A})$. We summarize the typical parameters in Table \ref{TABLESimu}, and all the simulations are configured according to the table unless stated otherwise. In this paper, the carrier frequency is $60$ GHz, the number of effective scattering paths is $L=5$, while other channel parameters are specified according to \cite{gao2016energy}. The antenna spacing is set as $d = \lambda$.

\begin{figure}
\center{\includegraphics[width=0.85\linewidth]{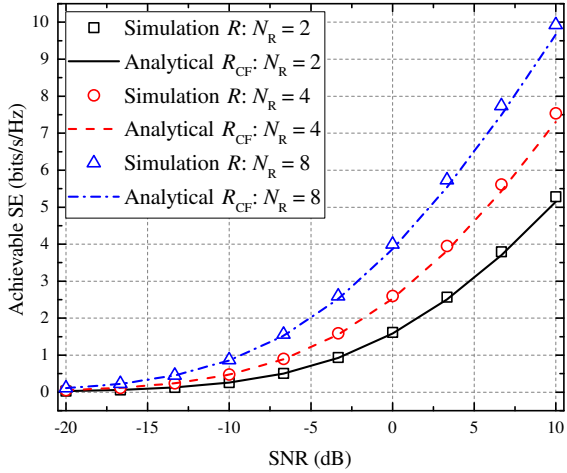}}
\caption{Simulated SE $R$ and analytical SE $R_\text{CF}$ averaged over $5,000$ random channel realizations with $N_\text{R} \in \{2, 4, 8\}$. Other parameters are specified according to Table \ref{TABLESimu}.}
\label{Fig_BoundTightness}
\end{figure}

In Fig.\ref{Fig_BoundTightness} we present the simulated SE $R(\mathbf{H}, \mathbf{A})$ and the analytical SE $R_\text{CF}(\mathbf{H}, \mathbf{A})$ averaged over $5, 000$ random channel realizations in conjunction with $N_\text{R} \in \{2, 4, 8\}$. Note that here we assume a trivial analog precoder, i.e. $\mathbf{A} = \mathbf{I}_{N_\text{T}} / \sqrt{N_\text{K}}$. It can then be seen that the proposed analytical expression $R_\text{CF}(\mathbf{H}, \mathbf{A})$ exhibits a favorable approximation accuracy to the simulation results. Hence we will use $R_\text{CF}(\mathbf{H}, \mathbf{A})$ as a low-complexity cost function to design the analog precoder $\mathbf{A}$.

\section{Proposed Analog Precoder Design}
The approximation (\ref{RLB}) provides a low-complexity alternative for evaluating the achievable rate, which can be harnessed to design the analog precoder $\mathbf{A}$. The optimal analog precoder can be obtained by solving the following problem:
\begin{equation}
  \mathbf{A}^\text{opt} = \displaystyle \arg \max_{\mathbf{A}} R_\text{CF} (\mathbf{H}, \mathbf{A}).
  \label{OptProblem0}
\end{equation}

Due to the complicated formulation of $R_\text{CF} (\mathbf{H}, \mathbf{A})$ in (\ref{RLB}), a closed-form solution to (\ref{OptProblem0}) is not directly accessible. Therefore we seek to derive the conjugate gradient of $R_\text{CF}(\mathbf{H}, \mathbf{A})$ with respect to $\mathbf{A}$, i.e. $\nabla_{\mathbf{A}^*} R_\text{CF}$. By applying the following denotations ($1 \le m, n \le M$):
\begin{equation}
\arraycolsep=1.0pt\def\arraystretch{1.3}
  \begin{array}{l}
  \displaystyle \mathbf{D}_m \triangleq \mathbf{C}_m \mathbf{C}_m^H, \,\, \rho_\text{S} \triangleq \frac{\rho}{\sigma_\text{N}^2 N_\text{S}} \\
  \displaystyle \bm\Gamma_{mn} \triangleq \mathbf{HA}\left( \mathbf{D}_m + \mathbf{D}_n \right) \mathbf{A}^H \mathbf{H}^H,
  \end{array}
  \label{bmGamma_Def}
\end{equation}
$\nabla_{\mathbf{A}^*} R_\text{CF}$ can thus be derived as in Equation (\ref{RLB_ConjGrdnt}). In order that $\mathbf{A}^\text{LO}$ is a local optimum of the optimization of (\ref{OptProblem0}), it is thus required that:
\begin{equation}
  \text{angle} \left[ \text{diag}\left( \mathbf{A}^\text{LO} \right) \right] = \text{angle} \left[ \text{diag}\left( \Delta \mathbf{A}^\text{LO} \right) \right],
  \label{LO_Criteria}
\end{equation}
where $\Delta \mathbf{A} \triangleq \nabla_{\mathbf{A}^*}R_\text{CF}$, $\text{angle}(\cdot)$ denotes the phase vector, and $\text{diag}(\cdot)$ represents the diagonal components of a matrix. The requirement of (\ref{LO_Criteria}) follows from the process of solving (\ref{OptProblem0}) via gradient ascent method. As a matter of fact, if a specific $\mathbf{A}$ satisfies (\ref{LO_Criteria}), then the gradient ascent operation imposes no impact on the phase vector of $\text{diag}(\mathbf{A})$, which leads to local convergence and $\mathbf{A}$ is thus a local optimum.

\begin{figure*}[ht!]
    \begin{equation}
    \nabla_{\mathbf{A}^*} R_\text{CF}\left( \mathbf{H}, \mathbf{A} \right) = \displaystyle \frac{\rho_\text{S} \log_2 e}{2M} \sum_{m=1}^M \frac{\sum_{t=1}^M \left| \mathbf{I}_{N_\text{R}} + \frac{\rho_\text{S}}{2} \bm\Gamma_{mt} \right|^{-1} \mathbf{H}^H \left( \mathbf{I}_{N_\text{R}} + \frac{\rho_\text{S}}{2} \bm\Gamma_{mt} \right)^{-1} \mathbf{HA}\left( \mathbf{D}_m + \mathbf{D}_t \right) }{\sum_{n=1}^M \left| \mathbf{I}_{N_\text{R}} + \frac{\rho_\text{S}}{2} \bm\Gamma_{nt} \right|^{-1}}.
    \label{RLB_ConjGrdnt}
    \end{equation}
\hrulefill
\end{figure*}

\begin{algorithm}[h]
 \caption{Maximizing $R_\text{CF}$ Over the Analog Precoder}
 \label{alg:Algorithm1}
 \begin{algorithmic}[1]

 \STATE \textit{Initialization}: Generate a feasible solution $\mathbf{A} = \text{diag}(\mathbf{a})$. Set $t = 0$ and given a maximum iteration number $t_\text{max}$.

 \STATE \textit{Calculate gradient}: Based on (\ref{RLB_ConjGrdnt}), calculate the conjugate gradient $\Delta \mathbf{A} = \nabla_{\mathbf{A}^*}R_\text{CF}$, and let $\Delta \mathbf{a} = \text{diag}\left( \Delta \mathbf{A} \right)$. \label{alg1:SearchDirection}

 \STATE \textit{Update}: Let $\mathbf{A} = \text{diag}\left[ e^{j \text{angle} \left( \Delta \mathbf{a} \right)}\right]$. Go to Step \ref{alg1:SearchDirection} until $t = t_\text{max}$.

 \end{algorithmic}
\end{algorithm}

Based on (\ref{LO_Criteria}), we propose our iterative algorithm to solve (\ref{OptProblem0}) in Algorithm \ref{alg:Algorithm1}. However, due to the $M^2$ matrices' conversions required by (\ref{RLB_ConjGrdnt}), the computational complexity of Algorithm \ref{alg:Algorithm1} could be massive when $M$ is large. Hence we seek to simplify the gradient calculation of (\ref{RLB_ConjGrdnt}) in the region of high SNR. More specifically, by exploiting the Woodbury matrix identity \cite{xzhang2004zhang}, the following approximation can be obtained when $m \ne t$:
\begin{equation}
  \arraycolsep=1.0pt\def\arraystretch{1.3}
  \begin{array}{rcl}
  && \displaystyle \left( \mathbf{I}_{N_\text{R}} + \frac{\rho_\text{S}}{2} \bm\Gamma_{mt} \right)^{-1} \\
  &=& \displaystyle \mathbf{I}_{N_\text{R}} - \frac{\rho_\text{S}}{2} \mathbf{HA} \mathbf{Q}_{mt} \left( \mathbf{I}_{2N_\text{RF}} + \frac{\rho_\text{S}}{2} \bm\Pi_{mt}\right)^{-1} \mathbf{Q}_{mt}^H \mathbf{A}^H \mathbf{H}^H \\
  &\overset{\text{(a)}}{\approx}& \displaystyle \mathbf{I}_{N_\text{R}} - \mathbf{HA} \mathbf{Q}_{mt} \bm\Pi_{mt}^{-1} \mathbf{Q}_{mt}^H \mathbf{A}^H \mathbf{H}^H,
  \end{array}
  \label{WoodburyApplication}
\end{equation}
where (a) is obtained at an asymptotically high SNR, i.e. $\rho_\text{S}/2 \gg 1$, and the following denotations are used:
\begin{equation}
  \displaystyle \mathbf{Q}_{mt} \triangleq \left[\mathbf{C}_m, \mathbf{C}_t\right], \,\, \bm\Pi_{mt} \triangleq \mathbf{Q}_{mt}^H \mathbf{A}^H \mathbf{H}^H \mathbf{HA} \mathbf{Q}_{mt}.
\end{equation}

Note that the approximation (\ref{WoodburyApplication}) only holds when $m \ne t$, as $\bm\Pi_{mt}$ is non-invertible when $m = t$. Therefore, the following approximation can be obtained by applying the high-SNR approximation of (\ref{WoodburyApplication}), when $m \ne t$:
\begin{equation}
  \displaystyle \mathbf{H}^H \left(\mathbf{I}_{N_\text{R}} + \frac{\rho_\text{S}}{2} \bm\Gamma_{mt}  \right)^{-1} \mathbf{HA} \left(\mathbf{D}_m + \mathbf{D}_t\right) \approx \mathbf{0}.
  \label{HighSNR_Approximation0}
\end{equation}

Moreover, by applying (\ref{WoodburyApplication}) and exploiting the determinant's property, i.e. $\left| \mathbf{I} + \mathbf{XY} \right| = \left| \mathbf{I} + \mathbf{YX} \right|$, the following approximation can also be obtained at high SNR when $m \ne t$:
\begin{equation}
  \arraycolsep=1.0pt\def\arraystretch{1.3}
  \begin{array}{rcl}
  && \displaystyle \left| \mathbf{I}_{N_\text{R}} + \frac{\rho_\text{S}}{2} \bm\Gamma_{mt} \right|^{-1} \\
  &\approx& \displaystyle \left| \mathbf{I}_{N_\text{R}} - \mathbf{HA} \mathbf{Q}_{mt} \bm\Pi_{mt}^{-1} \mathbf{Q}_{mt}^H \mathbf{A}^H \mathbf{H}^H \right| = 0.
  \end{array}
  \label{HighSNR_Approximation1}
\end{equation}

Based on (\ref{HighSNR_Approximation0}) and (\ref{HighSNR_Approximation1}), (\ref{RLB_ConjGrdnt}) can thus be simplified as follows when a high SNR is invoked:
\begin{equation}
  \arraycolsep=1.0pt\def\arraystretch{1.3}
  \begin{array}{rcl}
  && \nabla_{\mathbf{A}^*}R_\text{CF} \\
  &\approx& \displaystyle \frac{\rho_\text{S} \log_2 e}{M} \sum_{m=1}^M \mathbf{H}^H \left( \mathbf{I}_{N_\text{R}} + \frac{\rho_\text{S}}{2} \bm\Gamma_{mm} \right)^{-1} \mathbf{HA} \mathbf{D}_m,
  \end{array}
  \label{HighSNR_Approximation2}
\end{equation}
where $\bm\Gamma_{mm}$ has been given in (\ref{bmGamma_Def}). Again, by applying Woodbury matrix identity, we have:
\begin{equation}
  \arraycolsep=1.0pt\def\arraystretch{1.6}
  \begin{array}{rcl}
  && \displaystyle \mathbf{H}^H \left( \mathbf{I}_{N_\text{R}} + \frac{\rho_\text{S}}{2} \bm\Gamma_{mm} \right)^{-1} \mathbf{HA} \mathbf{D}_m \\
  &=& \displaystyle \mathbf{H}^H \mathbf{HAC}_m \left[\mathbf{C}_m^H  - \rho_\text{S} \left(\mathbf{I}_{N_\text{RF}} + \text{...} \right.\right.\\
  && \displaystyle \left.\left. \rho_\text{S} \mathbf{C}_m^H \mathbf{A}^H \mathbf{H}^H \mathbf{HAC}_m \right)^{-1} \mathbf{C}_m^H \mathbf{A}^H \mathbf{H}^H \mathbf{HAD}_m \right] \\
  &\overset{\text{(a)}}{\approx}& \displaystyle \rho_\text{S}^{-1} \mathbf{H}^H \mathbf{HAC}_m \left( \mathbf{C}_m^H \mathbf{A}^H \mathbf{H}^H \mathbf{HAC}_m \right)^{-1},
  \end{array}
  \label{HighSNR_Approximation3}
\end{equation}
where (a) is yielded by applying the following approximation at an asymptotically high SNR:
\begin{equation}
  \arraycolsep=1.0pt\def\arraystretch{1.6}
  \begin{array}{rcl}
  && \displaystyle \left( \mathbf{I}_{N_\text{R}} + \rho_\text{S} \mathbf{C}_m^H \mathbf{A}^H \mathbf{H}^H \mathbf{HAC}_m \right)^{-1} \\
  &\approx& \displaystyle \left[\mathbf{I}_{N_\text{RF}} - \rho_\text{S}^{-1} \left( \mathbf{C}_m^H \mathbf{A}^H \mathbf{H}^H \mathbf{HAC}_m \right)^{-1} \right] \times \text{...} \\
  && \displaystyle \left(\mathbf{C}_m^H \mathbf{A}^H \mathbf{H}^H \mathbf{HAC}_m \right)^{-1} \rho_\text{S}^{-1}.
  \end{array}
  \label{HighSNR_Approximation4}
\end{equation}

Finally, substituting (\ref{HighSNR_Approximation3}) into (\ref{HighSNR_Approximation2}) yields the following low-complexity approximation of $\nabla_{\mathbf{A}^*} R_\text{CF}$:
\begin{equation}
  \arraycolsep=1.0pt\def\arraystretch{1.6}
  \begin{array}{rcl}
  && \displaystyle \nabla_{\mathbf{A}^*} R_\text{CF} \\
  &\approx& \displaystyle \frac{\log_2 e}{M} \sum_{m=1}^M \mathbf{H}^H \mathbf{H} \mathbf{AC}_m \left( \mathbf{C}_m^H \mathbf{A}^H \mathbf{H}^H \mathbf{HAC}_m \right)^{-1}.
  \end{array}
  \label{RLB_ConjApprox}
\end{equation}

Comparing (\ref{RLB_ConjApprox}) with (\ref{RLB_ConjGrdnt}), it can be seen that (\ref{RLB_ConjGrdnt}) requires $M^2$ inversions of $(N_\text{R} \times N_\text{R})$-dimensional matrices, while (\ref{RLB_ConjApprox}) only requires $M$ inversions of $(N_\text{RF} \times N_\text{RF})$-dimensional matrices. Therefore, by exploiting (\ref{RLB_ConjApprox}) as the alternative gradient expression, the complexity order can be reduced from $\mathcal{O}(M^2 N_\text{R}^3)$ to $\mathcal{O}(M N_\text{RF}^3)$. For convenience, in this paper, Algorithm \ref{alg:Algorithm1} using the gradient expression of (\ref{RLB_ConjGrdnt}) is referred to as the \textit{full-complexity algorithm}, while Algorithm \ref{alg:Algorithm1} using the gradient expression of (\ref{RLB_ConjApprox}) is referred to as the \textit{reduced-complexity algorithm}. However, it is worth noting that the matrix inversion in (\ref{RLB_ConjApprox}) requires that the channel rank is not smaller than $N_\text{RF}$, i.e. $\text{rank}(\mathbf{H}) \ge N_\text{RF}$, otherwise the reduced-complexity algorithm would not be applicable.

\section{Simulation Results}
In this section we present the achievable SE of the proposed scheme via numerical simulations. The precoding scheme of \cite{ishikawa2016GSM} corresponds to the case of $\mathbf{A} = \mathbf{I}_{N_\text{T}} / \sqrt{N_\text{K}}$, while the scheme of \cite{liu2016LOS} corresponds to the case of $N_\text{M} = N_\text{T}$ and $N_\text{K} = 1$ with no precoding. The performance presented in this section are all obtained based on the true SE expression $R(\mathbf{H}, \mathbf{A})$. Other parameters are the same as Section \uppercase\expandafter{\romannumeral2}-C. Note that we set the number of effective scattering paths to $L = 5$ so that $\text{rank}(\mathbf{H}) \ge N_\text{RF}$ holds, as required by (\ref{RLB_ConjApprox}).

\subsection{Parameter Selection}
We commence by discussing the parameter selection in our proposed system. More specifically, we focus on the selection of $(N_\text{K}, N_\text{M})$ under the constraint of $N_\text{K} N_\text{M} = N_\text{T}$, while keeping $N_\text{T}$ and $N_\text{RF}$ invariant. Intuitively, increasing $N_\text{K}$ leads to reducing $N_\text{M}$ and the corresponding spatial multiplexing gain brought by GenSM, while it also leads to a larger antenna group and therefore enhances the array gain provided by the analog precoder. Hence $N_\text{K}$ is the key to a scalable tradeoff between multiplexing gain and array gain.

Given the favorable approximation accuracy provided by the proposed closed-form expression $R_\text{CF}(\mathbf{H}, \mathbf{A})$, as demonstrated by Fig.\ref{Fig_BoundTightness}, it is thus reasonable to use $R_\text{CF}(\mathbf{H}, \mathbf{A})$, instead of $R(\mathbf{H}, \mathbf{A})$, as a low-complexity metric to evaluate the parameter optimality. More specifically, we propose to optimize the analog precoder $\mathbf{A}^*$ by performing Algorithm \ref{alg:Algorithm1} for each $(N_\text{K}, N_\text{M})$ pair, and then select the optimal parameters so that $R_\text{CF}(\mathbf{H}, \mathbf{A}^*)$ is maximized. In Table \ref{TABLEDifferentParams}, we present the optimal $(N_\text{K}, N_\text{M})$ selection yielded by various system parameters in conjunction with the optimized analog precoding. In general, a larger $N_\text{M}$ is desired, when a higher SNR value or a larger $N_\text{R}$ is applied. This finding motivates us to apply a larger value of $N_\text{M}$ when the receiver is in a sufficiently good condition (not-so-small $N_\text{R}$ or SNR) to harness the GenSM gain.

\begin{table}
\small
\caption{Optimal $(N_\text{K}, N_\text{M})$ Yielded by Various $N_\text{R}$ and SNR With Optimized Precoding. Other Parameters Are $N_\text{T} = 8$ and $N_\text{RF} = 1$}
\centering
\renewcommand\arraystretch{1.2}
\begin{tabular}{c|c|c|c|c|c}
\hline\hline
$N_\text{R}$ & 4 & 6 & \multicolumn{3}{c}{8} \\\hline
SNR (dB) & 10 & 10 & 3 & 6 & 10 \\\hline
Optimal $(N_\text{K}, N_\text{M})$ & (8, 1) & (2, 4) & (8, 1) & (2, 4) & (1, 8) \\\hline
\hline
\end{tabular}
\label{TABLEDifferentParams}
\end{table}

\subsection{SE Comparison}

\begin{figure}
\center{\includegraphics[width=0.80\linewidth]{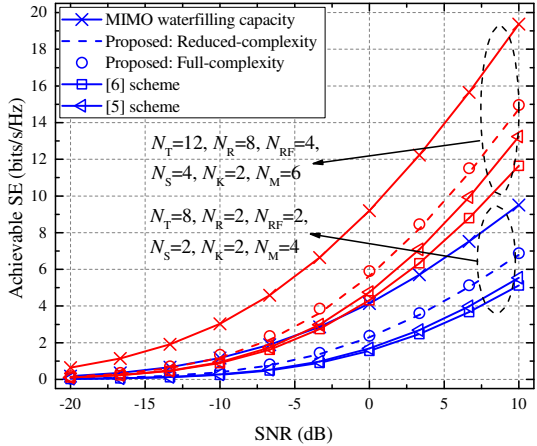}}
\caption{SE performance achieved by different schemes associated with various parameters. Other parameters are specified according to Table \ref{TABLESimu}.}
\label{Fig_PerformanceComparison}
\end{figure}

In Fig.\ref{Fig_PerformanceComparison}, the SE performance yielded by various schemes and the MIMO waterfilling capacity are depicted. It is worth noting that, due to the high cost of RF chains, the MIMO waterfilling capacity should be achieved with the same $N_\text{RF}$ as our proposed scheme for fairness.

According to Fig.\ref{Fig_PerformanceComparison}, the performance of the proposed reduced-complexity algorithm is almost the same as that of the full-complexity counterpart, which substantiates its better performance-complexity tradeoff. Moreover, it is also shown that the proposed scheme outperforms the schemes of \cite{ishikawa2016GSM} and \cite{liu2016LOS}. The reason is that the schemes of \cite{ishikawa2016GSM} and \cite{liu2016LOS} fail to sufficiently exploit the CSI at the transmitter. Besides, the schemes of \cite{ishikawa2016GSM} and \cite{liu2016LOS} are designed for LoS channels, which are different from the Saleh-Valenzuela channel model utilized in this paper and hence leads to performance penalty.

Lastly, it is seen that the proposed scheme is outperformed by the MIMO waterfilling precoder. The reason is that our scheme solely considers the application of analog precoder, while a \textit{digital precoder} is also essential for achieving a near-optimal SE performance.

\subsection{BER Comparison}
\begin{figure}
\center{\includegraphics[width=0.80\linewidth]{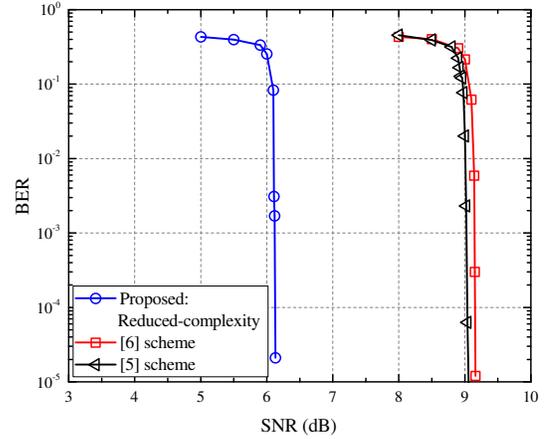}}
\caption{BER performance of various schemes. The proposed scheme and the scheme of \cite{ishikawa2016GSM} is associated with $N_\text{S} = 2$, $M = 4$ and QPSK, while the scheme of \cite{liu2016LOS} is associated with $N_\text{S} = 2$, $M=16$ and BPSK. The LDPC coding rate is $1/2$ and the coding block length is $32,400$ with $16,000$ random channel realizations.}
\label{Fig_BERComparison}
\end{figure}

Finally, we depict the BER performance of various schemes in Fig.\ref{Fig_BERComparison}. The performance of the proposed scheme and \cite{ishikawa2016GSM} is yielded by $N_\text{K}=2$, $N_\text{M} = 4$, $N_\text{S} = N_\text{RF} = 2$, $N_\text{R}=2$ and BPSK (which leads to $6$ bits per channel use), while the performance of \cite{liu2016LOS} is yielded by $N_\text{K}=1$, $N_\text{T} = N_\text{M} = 8$, $N_\text{S} = N_\text{RF} = 2$, $N_\text{R} = 2$ and BPSK (which also leads to $6$ bits per channel use). The LDPC coding rate is $1/2$ with coding block length $32,400$ (MATLAB functions $\mathsf{comm.LDPCEncoder}$ and $\mathsf{comm.LDPCDecoder}$ are utilized), hence the normalized throughput is $3$ bits per channel use for all the schemes. It can be readily seen that the proposed scheme outperforms other schemes by approximately $3$ dB, which substantiates the superior performance of the proposed system.

\section{Conclusions}
This paper investigated the analog precoder design for GenSM-aided mmWave MIMOs. A closed-form approximation was proposed to quantify the achievable rate of the proposed scheme. Based on the closed-form expression, iterative algorithms were proposed to design the analog precoder. Lastly, numerical simulations were provided to demonstrate the superior performance of the proposed scheme.

\end{document}